\documentclass[fleqn,12pt,twoside]{article}
\usepackage{espcrc1}
\usepackage{graphicx}
\usepackage{epsfig}
\usepackage[figuresright]{rotating}

\newcommand{\AmS}{{\protect\the\textfont2
  A\kern-.1667em\lower.5ex\hbox{M}\kern-.125emS}}
\title{Results on Identified Charged Hadrons from the PHENIX Experiment at RHIC}
\author{T. Chujo\address[BNL]{Brookhaven National Laboratory, Upton, NY 11973-5000, USA} 
for the PHENIX Collaboration.
}
       
\begin{document}
\maketitle

\begin{abstract}
Recent results on identified hadrons from the PHENIX experiment in Au+Au 
collisions at mid-rapidity at $\sqrt{s_{NN}}$ = 200 GeV are presented. 
The centrality dependence of transverse momentum distributions and particle 
ratios for identified charged hadrons are studied. 
\end{abstract}

\section{Introduction and Experimental Setup}
The physics motivation of the ultra-relativistic heavy-ion program at 
the Relativistic Heavy Ion Collider (RHIC) is to study nuclear matter 
at extremely high temperature and energy density with the hope to reach 
a new form of matter called the quark gluon plasma (QGP). Among the various 
probes of the QGP state, hadrons carry important information about the 
collision dynamics along with the spatial and temporal evolution of the 
system from the early stage of the collisions to the final state interactions.

For studies of hadron physics at RHIC, the PHENIX experiment~\cite{NIM} 
demonstrates good capability for particle identification (PID) for both 
charged hadrons ($\pi^{\pm}$, $K^{\pm}$, $p$, $\overline{p}$, $d$ and 
$\overline{d}$) and neutral pions over a broad momentum range. The charged 
hadrons can be identified with time-of-flight measurements in two different 
detectors: (1) a high resolution Time-of-Flight wall (TOF) and, 
(2) an electro-magnetic calorimeter (EMC), in conjunction with the tracking 
system in the PHENIX central arm spectrometers and the beam counter, which 
provides the start timing and the event vertex determination. The tracking 
system in the central arm consists of drift chambers (DC), three layers of 
pad chambers (PC), and time expansion chambers (TEC). The PHENIX central arms 
cover $|\eta|<0.35$ in pseudo-rapidity, and cover $\pi/4$ with the TOF and 
$3\pi/4$ by EMC in azimuth. The $\pi/K$ and $K/p$ separation can be achieved 
up to 2 and 4 GeV/$c$ in $p_{T}$, respectively, using the TOF detector, which 
has a 120 ps timing resolution. Neutral pions are identified with the EMC via 
the $\pi^0 \rightarrow \gamma\gamma$ decay channel up to 10 GeV/$c$ in $p_{T}$ 
using the full statistics taken during Run II at RHIC in Au+Au collisions~\cite{David}.  

\section{Experimental Results}

\begin{figure}[t]
\begin{center}
\includegraphics[width=12cm]{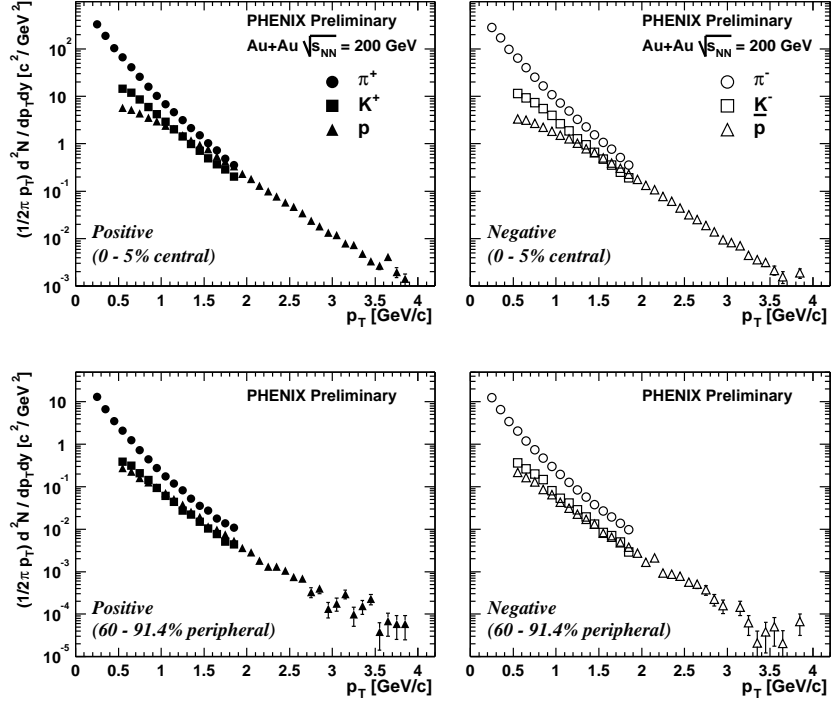}
\caption{Transverse momentum distributions for pions (circles), 
kaons (squares) and $p$, $\overline{p}$ (triangles) in the 0--5\% 
most central events (upper panels) and 60--91.4\% most
peripheral events (lower panels) at $\sqrt{s_{NN}}$ = 200 GeV 
in Au+Au collisions. The left panels show positive particles 
and the right panels show negative particles . The error bars are 
statistical only.}
\label{fig:Spectra_cent}
\end{center}
\end{figure}

\begin{figure}[h]
\begin{center}
\includegraphics[width=12cm]{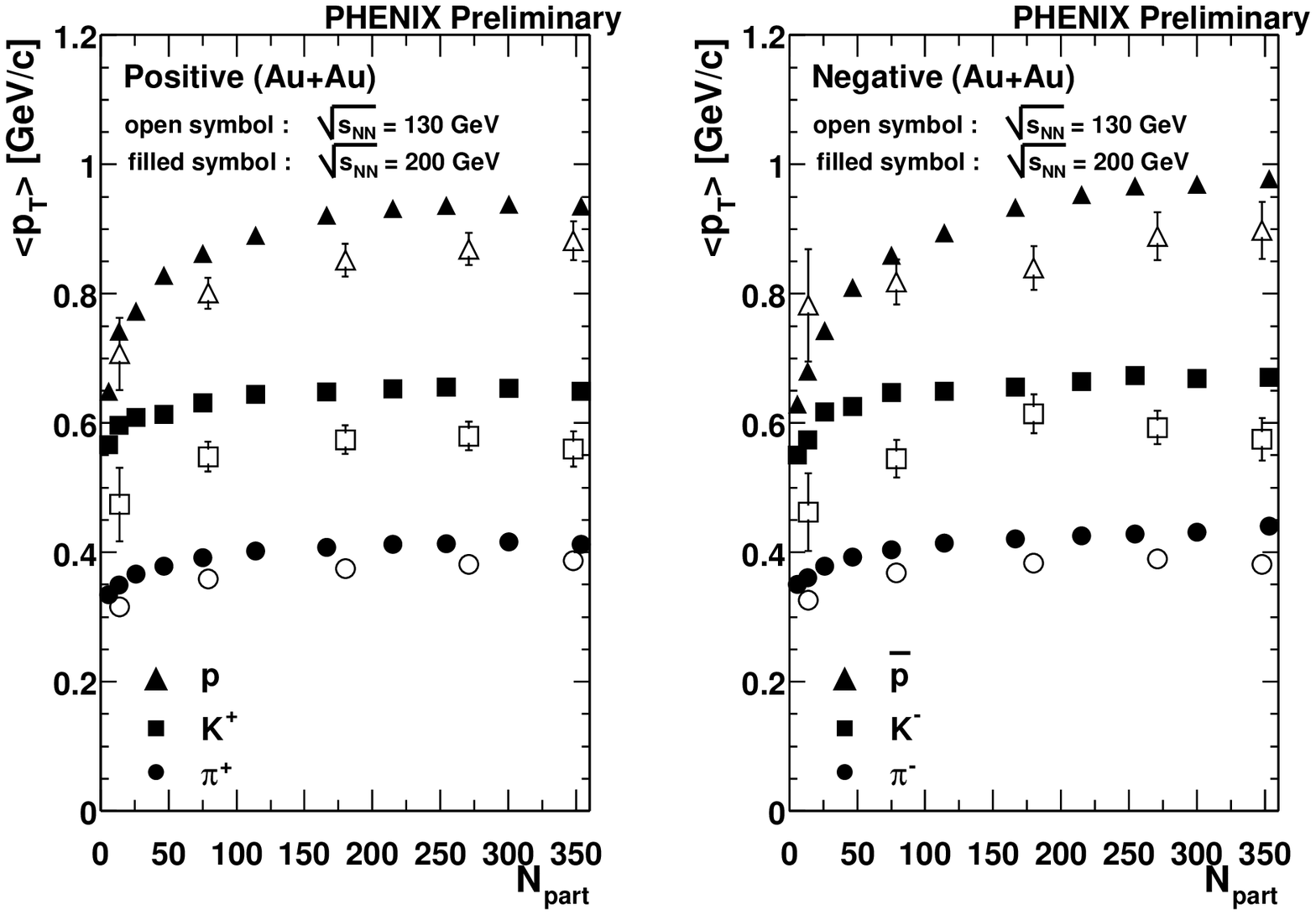}
\caption{Mean transverse momentum for identified charged
hadrons as a function of $N_{part}$ for protons and anti-protons 
(triangles), kaons (squares) and pions (circles). The left panel
shows positive particles and right panel shows negative particles. 
The open symbols indicate the data for 130 GeV~\cite{PPG006} while the 
filled symbols indicate the data for 200 GeV in Au+Au collisions.}
\label{fig:meanPt}
\end{center}
\end{figure}

We have measured the transverse momentum distributions for $\pi^{\pm}$, 
$K^{\pm}$, $p$ and $\overline{p}$ at mid-rapidity in Au+Au collisions 
at $\sqrt{s_{NN}}$ = 200 GeV over a broad momentum range over various 
centrality selections. To identify the charged particles, the high 
resolution TOF counter is used in this analysis. We use about 4 million 
minimum bias events. The data are classified into 11 centrality bins
expressed in percent of the total inelastic cross section. 
The spectra for each particle species are corrected for geometrical 
acceptance, decay in flight, multiple scattering, and tracking efficiency 
using a single particle Monte Carlo simulation. A multiplicity-dependent 
track reconstruction efficiency is also determined and applied by embedding 
simulated tracks into real events.

The upper two panels in Figure~\ref{fig:Spectra_cent} show the $p_T$ 
distributions for identified hadrons in the most central events (0--5\%) 
in 200 GeV Au+Au collisions for positive (left) and negative (right)
particles. The lower two panels show the most peripheral events (60--91.4\%).
In each panel, the data are presented up to 1.8 GeV/$c$ for charged 
pions and kaons, and 3.8 GeV/$c$ for $p$ and $\overline{p}$. In the 
low $p_T$ region of the most central events, the data indicate that the 
inverse slope increases with the particle mass. Also, the shape of the $p$ 
and $\overline{p}$  spectra have a shoulder-arm shape while the pion spectra 
have a concave shape. On the other hand, in the most peripheral events, the 
spectra are almost parallel to each other. This mass dependence of the slopes 
and shapes of the spectra for protons in the central events can 
be explained by a radial flow picture. At around 2.0 GeV/$c$  
in central events, the proton yield becomes comparable to the pion yield.
A similar behavior is also seen for negatively charged particles. 

In order to quantify the observed mass dependence of the slopes,
the mean transverse momenta $\langle p_{T} \rangle$ are extracted 
for 11 centrality selections and for each particle species.
The $p_T$ spectra are extrapolated to below and above the measured range
using a power law function for pions, an $m_{T}$ exponential function for 
kaons, and a Boltzmann function for $p$ and $\overline{p}$. 
In Figure~\ref{fig:meanPt}, the centrality dependence of 
$\langle p_{T} \rangle$ for identified charged hadrons are shown
together with the 130 GeV data points~\cite{PPG006}. 
In both the 200 GeV and 130 GeV data, $\langle p_{T} \rangle$ for
all particle species increases from the most peripheral to the most 
central events and also increases with particle mass. The dependence of 
$\langle p_{T} \rangle$ on particle mass suggests the existence of 
a collective hydrodynamical expansion. The dependence on the number 
of participant nucleons ($N_{part}$) calculated using a Glauber 
model~\cite{Glauber} may be due to an increasing 
radial expansion from peripheral to central events.

\begin{figure}
\begin{center}
\includegraphics[width=14cm]{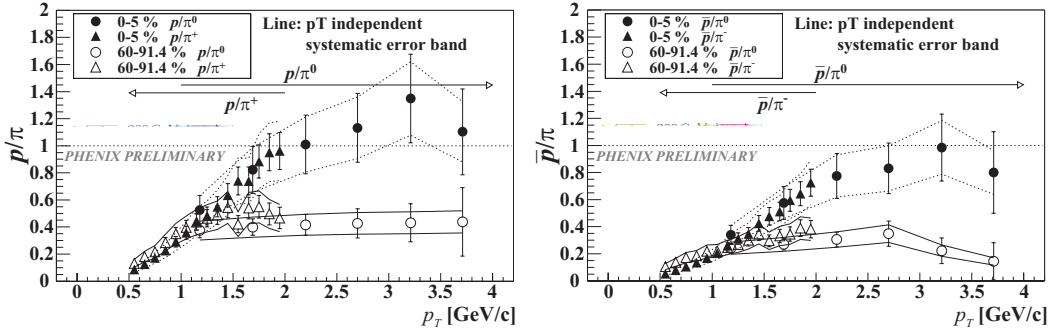}
\caption{ $p/\pi$ (left) and $\overline{p}/\pi$ (right) ratios 
as a function of $p_T$ for the 0--5 \% and 60--91.4\% centrality selections. 
Lines along the data points show the $p_T$-independent systematic error bands.
The error bars show the statistical and $p_T$-dependent systematic errors, 
summed in quadrature.}
\label{fig:ppi_ratio}
\end{center}
\end{figure}

Particle composition at high $p_T$ is also interesting in order to understand 
baryon production and transport, system evolution, and the interplay 
between soft processes and jet quenching in hard processes. In 
Figure~\ref{fig:ppi_ratio},
the $p/\pi$ and $\overline{p}/\pi$ ratios are shown. For the denominator 
of the ratio $p/\pi$ and $\overline{p}/\pi$, we use
two independent measurements with different subsystems: (1) $p_T$ 
spectra for $\pi^{\pm}$ up to 2 GeV/$c$ identified using the TOF, (2) neutral 
pions results from 1 GeV/$c$ to 3.8 GeV/$c$ measured using the EMC. 
Both the $p/\pi$ and $\overline{p}/\pi$ ratios have a clear centrality 
dependence. The data shows that these ratios in central collisions reach
unity at a $p_{T}$ of 2$\sim$3 GeV/$c$, while they saturate at around 
$p_{T}$ of 0.3 -- 0.4 in peripheral collisions. This observed behavior in 
central events may be attributed to the composition of two effects including 
a larger flow effect for protons (and $\overline{p}$) compared to pions, and 
a pion suppression effect at high $p_{T}$~\cite{PPG003}.

\section{Summary}
We present results on identified hadrons in Au+Au collisions at 
$\sqrt{s_{NN}}$ = 200 GeV at mid-rapidity over different centrality 
selections from the PHENIX experiment. The transverse momentum distributions 
for $\pi^{\pm}$, $K^{\pm}$, $p$, $\overline{p}$, $d$, and $\overline{d}$ 
are measured and we observe a mass dependence of the slopes and
shapes of the identified charged spectra in central events. 
However, they are almost parallel to each other in the most peripheral events. 
Also, the mean transverse momentum for all particle species increases from 
peripheral to central events, and with particle mass. The ratios $p/\pi$ and 
$\overline{p}/\pi$ up to 3.8 GeV/$c$ are also measured by using the combined 
results for charged and neutral pions. In central collisions, the ratio reaches 
unity at a $p_{T}$ of 2$\sim$ 3 GeV/$c$ and saturates at a $p_{T}$ 
around 0.3 -- 0.4 GeV/$c$ in peripheral collisions.

\end{document}